\documentclass[sigconf,nonacm]{acmart}
\usepackage{enumitem}
\usepackage{graphicx}
\usepackage{framed}
\usepackage{algorithm}
\usepackage{algpseudocode} 
\usepackage{multirow}
\usepackage{pifont}
\usepackage{amsmath}
\usepackage[skip=0pt]{caption}
\usepackage{fvextra}
\usepackage{listings}
\usepackage{xcolor}
\usepackage{setspace}
\usepackage[most]{tcolorbox}
\copyrightyear{2026}
\acmYear{2026}
\setcopyright{cc}
\setcctype{by}
\acmConference[WWW '26]{Proceedings of the ACM Web Conference 2026}{April 13--17, 2026}{Dubai, United Arab Emirates}
\acmBooktitle{Proceedings of the ACM Web Conference 2026 (WWW '26), April 13--17, 2026, Dubai, United Arab Emirates}
\acmPrice{}
\acmDOI{10.1145/3774904.3792229}
\acmISBN{979-8-4007-2307-0/2026/04}
\begin{document}

\title{DMAP: Human-Aligned Structural Document Map for Multimodal Document Understanding}

\author{ShunLiang Fu}
\email{forlorin@njust.edu.cn}
\affiliation{%
  \institution{Nanjing University of Science and Technology}
  \city{Nanjing}
  \state{Jiangsu}
  \country{China}
}
\author{Yanxin Zhang}
\email{yzhang2879@wisc.edu}
\affiliation{%
  \institution{University of Wisconsin--Madison}
  \city{Madison}
  \state{Wisconsin}
  \country{United States}
}

\author{Yixin Xiang}
\email{elephantoh@qq.com}
\affiliation{%
  \institution{Nanjing University of Science and Technology}
  \city{Nanjing}
  \state{Jiangsu}
  \country{China}
}

\author{Xiaoyu Du}
\email{duxy@njust.edu.cn}
\affiliation{%
  \institution{Nanjing University of Science and Technology}
  \city{Nanjing}
  \state{Jiangsu}
  \country{China}
}
\authornote{Corresponding author.}

\author{Jinhui Tang}
\email{tangjh@njfu.edu.cn}
\affiliation{%
  \institution{Nanjing Forestry University}
  \city{Nanjing}
  \state{Jiangsu}
  \country{China}
}

\newcommand{\eg}{\textit{e.g.}}
\newcommand{\ie}{\textit{i.e.}}
\newcommand{\etc}{\textit{etc.}}

\newcommand{\SummaryName}{DMAP}
\hyphenation{data-sets}
\hyphenation{Mind-Map}
\hyphenation{Paper-Text}




\begin{abstract}
Existing multimodal document question-answering (QA) systems predominantly rely on flat semantic retrieval, representing documents as a set of disconnected text chunks and largely neglecting their intrinsic hierarchical and relational structures. Such flattening disrupts logical and spatial dependencies—such as section organization, figure-text correspondence, and cross-reference relations—that humans naturally exploit for comprehension. To address this limitation, we introduce a document-level structural \textbf{D}ocument \textbf{MAP}~(DMAP), which explicitly encodes both hierarchical organization and inter-element relationships within multimodal documents. Specifically, we design a Structured-Semantic Understanding Agent to construct DMAP by organizing textual content together with figures, tables, charts, \text{\etc} into a human-aligned hierarchical schema that captures both semantic and layout dependencies. Building upon this representation, a Reflective Reasoning Agent performs structure-aware and evidence-driven reasoning, dynamically assessing the sufficiency of retrieved context and iteratively refining answers through targeted interactions with DMAP. Extensive experiments on MMDocQA benchmarks demonstrate that DMAP yields document-specific structural representations aligned with human interpretive patterns, substantially enhancing retrieval precision, reasoning consistency, and multimodal comprehension over conventional RAG-based approaches. Code is available at https://github.com/Forlorin/DMAP

\end{abstract}

\maketitle

\section{Introduction}
Multimodal document understanding aims to extract and reason over textual and visual information—such as figures, tables, charts, and text—within documents~\cite{hanMDocAgentMultiModalMultiAgent2025,jainSimpleDocMultiModalDocument2025,liuStructuredAttentionMatters2025,dongBenchmarkingRetrievalAugmentedMultimomal2025}.
This task is challenging for two main reasons: a) raw documents are complex and heterogeneous, making it difficult to organize and interpret the information, and b) modeling the relationships among different elements to support reasoning is non-trivial.
Documents are designed for human consumption, containing hierarchical cues such as titles, sections, paragraphs, and cross-page references, as well as intra-page visual-text alignments.
Efficient understanding requires preserving these structural cues while enabling reasoning over both content and layout.

\begin{figure}[ht]
\centering
\includegraphics[width=0.95\linewidth]{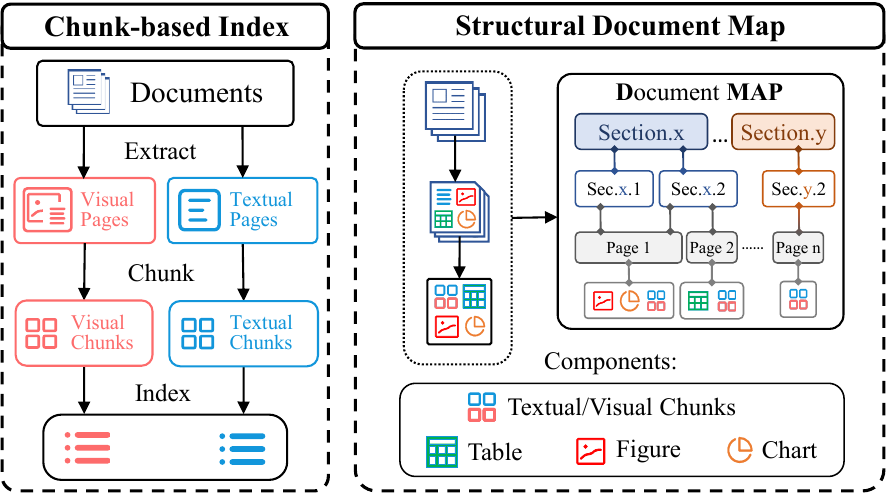}
\caption{The utilities of document knowledge.}
\label{fig:contrast}
\end{figure}

A common solution for document question answering is \emph{Retrieval-Augmented Generation} (RAG), where documents are split into several chunks and mapped into a high-dimensional semantic vector space~\cite{choM3DocRAGMultimodalRetrieval2024,liAgenticRAGDeep2025,dongUnderstandWhatLLM2025,renLexiconEnhancedSelfSupervisedTraining2022,liBridgingGapAligning2025}.
Queries are then matched to chunks based on vector similarity.
While effective in capturing local semantic relevance, such methods destroy the document’s inherent structure.
Flattening causes the loss of hierarchical and relational cues, leading to broken causal relationships, inconsistent parallel reasoning, and failures in handling referential expressions (\eg, `Table X' or `Page X')~\cite{sarthiRAPTORRecursiveAbstractive2023,prockoGraphRetrievalAugmentedGeneration2024,wuMoLoRAGBootstrappingDocument2025,zhangLinkNERLinkingLocal2024}.
Recent approaches, such as MDocAgent~\cite{hanMDocAgentMultiModalMultiAgent2025}, attempt to leverage LLMs or LVLMs to refine retrieval outputs with the whole page layout, but they still neglect the natural structural organization of documents—an aspect so significant that humans intuitively exploit it to locate, contextualize, and reason about information.

As illustrated in Figure~\ref{fig:contrast}, the conventional manner is to build a chunk-based index by fragmenting both textual and visual pages into isolated chunks and indexing them semantically, thereby losing the structural integrity of the original document.
This flattening disrupts the hierarchical and relational dependencies among document elements—for example, the logical progression between paragraphs or the alignment between a figure and its corresponding caption.
Such relations are central to human understanding but are entirely absent from traditional chunk-based retrieval.
The inability to preserve structure often leads to incoherent reasoning paths, missing causal or comparative cues, and incorrect scope during retrieval.
To achieve human-aligned document understanding, it is therefore crucial to restore and exploit the inherent organizational structure of documents.

To bridge this gap, we introduce a human-aligned structural \textbf{D}ocument \textbf{MAP}~(DMAP), a document-level representation that explicitly encodes hierarchical organization and inter-element relationships.
As shown in Figure~\ref{fig:contrast}, DMAP serves as a unified knowledge map that captures both the semantic and spatial dependencies among textual, tabular, and visual components, thereby preserving the intrinsic structure of the document.
To construct DMAP for specific documents, we design a Structured-Semantic Understanding Agent~(SSUA), which analyzes the document’s multimodal content and organizes textual segments, figures, tables, and charts into a human-intuitive hierarchical layout.
This process transforms unstructured document inputs into a well-formed DMAP, where relationships such as section containment, figure-text alignment, and cross-page linkage are explicitly modeled.
In order to validate the effectiveness of DMAP, we further devise a Reflective Reasoning Agent (RRA) that incorporates DMAP for the multimodal question answering task.
Unlike conventional RAG-based generators that directly rely on retrieved chunks, RRA leverages DMAP as a structured knowledge base: it heuristically evaluates whether the current evidence is sufficient for accurate reasoning, and—when necessary—actively queries DMAP to locate supplementary information.
This iterative reflection process mimics how humans verify and refine their understanding, ensuring that the generated answers are both semantically grounded and structurally consistent with the document.

We evaluate DMAP on MMDocQA (multi-modal document question answering) benchmarks, demonstrating that it effectively preserves structural and relational cues and consistently improves multimodal question answering performance.

Our contributions are three-fold:
\begin{enumerate}[leftmargin=*]
\item We identify the limitations of existing structure-agnostic RAG methods and reveal the necessity of explicitly modeling document structures for faithful multimodal understanding.

\item We propose DMAP, a human-aligned hierarchical representation that integrates textual, tabular, and visual information, and design SSUA (Structured-Semantic Understanding Agent) to automatically construct DMAP from multimodal documents.

\item We further design RRA (Reasoning and Response Agent), which leverages DMAP to perform structure-aware retrieval and reasoning, producing coherent and reasoning-consistent answers.
\end{enumerate}

\section{Related Work}
\textbf{LLMs for Document Understanding.}
With the emergence of large language models (LLMs),
document understanding has made rapid progress.
However, early LLMs and their multimodal counterparts (MLLMs) offered limited context window lengths.
As a result, document understanding systems built on these models primarily targeted single-page or short documents,
limiting comprehensive, fine-grained understanding of long documents~\cite{mishraOCRVQAVisualQuestion2019,tanakaSlideVQADatasetDocument2023,pradeepHowDoesGenerative2023}.

Recently, with the rapid advancements in LLM capabilities,
numerous models have been introduced that support substantially longer context windows,
such as Qwen-VL~\cite{baiQwenVLVersatileVisionLanguage2023} and GPT-4o~\cite{openaiGPT4oSystemCard2024}.
This progress has made it feasible to directly process multi-page documents as input~\cite{liLongContextVs2024}.
At the same time, some studies have focused on developing multimodal models tailored specifically for DocQA,
targeting further extensions of supported context length and improved comprehension of multimodal information~
\cite{titoHierarchicalMultimodalTransformers2023,huMPLUGDocOwl15Unified2024, huangLayoutLMv3PretrainingDocument2022}.

Nevertheless, these approaches still face limitations.
Although the maximum context length of LLMs has been significantly extended,
processing long documents remains challenging due to constraints in context window size and computational resources.
Moreover, fine-grained and key pieces of information in a document can still be easily overlooked when embedded in overly long contexts.
Therefore, an important research challenge lies in enabling DocQA systems to both retrieve sufficient contextual information
and utilize it effectively for accurate and detailed comprehension.

\textbf{Retrieval-Augmented Generation.}
Retrieval-Augmented Generation (RAG) is an alternative approach to improving the performance of large language models,
complementing fine-tuning and prompt engineering\cite{gaoRetrievalAugmentedGenerationLarge2024,jinAPEERAutomaticPrompt2025}.
By integrating retrieval with generation,
RAG enables models to leverage information from external knowledge bases during the generation process,
thereby enhancing both the accuracy and richness of responses.

In traditional RAG systems for document-based question answering,
documents are typically indexed as a whole when constructing the knowledge base~\cite{zhangGenerativeRetrievalTerm2024}.
This method works well for relatively short documents;
however, for longer documents, it often results in the loss of fine-grained information~
\cite{yueInferenceScalingLongContext2024}.

To address this challenge, long documents are typically divided into multiple semantically coherent segments,
which are subsequently transformed into vector representations for dense
indexing~\cite{gaoRetrievalAugmentedGenerationLarge2024,xuLargeLanguageModels2024}.
The retrieval quality in such approaches is largely determined by the effectiveness of the vectorization model.
Early implementations used general-purpose pre-trained language models such as BERT and LLaMA for
vectorization~\cite{maFineTuningLLaMAMultiStage2024,zhuangPromptRepsPromptingLarge2024, dongUnsupervisedLargeLanguage2024}.
As the demand for higher-quality text embeddings increased,
specialized embedding models, such as \emph{FlagEmbedding}, were proposed~\cite{zhangRetrieveAnythingAugment2023}.
For document embedding, ColBERT~\cite{santhanamColBERTv2EffectiveEfficient2022} introduced a late-interaction mechanism,
enabling richer interactions between user queries and document chunks.
Dense retrieval techniques have also been extended to the visual domain~\cite{zhouVISTAVisualizedText2024},
exemplified by Colpali~\cite{faysseColPaliEfficientDocument2024}, a visual adaptation of ColBERT.
In addition to dense retrieval, traditional sparse retrieval methods have also been explored in RAG systems,
with several studies proposing hybrid retrieval paradigms that integrate both dense and sparse
methods~\cite{formalSPLADESparseLexical2021,lassanceSPLADEv3NewBaselines2024,chenSTAIRLearningSparse2023}.

LLMs have also been employed to enhance retrieval performance.
There are several ways in which LLMs can improve retrieval, including query rewriting, re-ranking, and task
decomposition~\cite{liAgenticRAGDeep2025,maQueryRewritingRetrievalAugmented2023, asaiSelfRAGLearningRetrieve2023, maHetGPTHarnessingPower2024,chenTourRankUtilizingLarge2025,hongLLMBSEnhancingLarge2025}.
For example, ControlRetriever~\cite{panControlRetrieverHarnessingPower2023}
introduces a parameter-isolated architecture that leverages natural language instructions to control dense retrievers,
thereby enabling a unified and task-agnostic approach across diverse retrieval tasks.
In parallel, Simple-Doc~\cite{jainSimpleDocMultiModalDocument2025} proposes a strategy for retrieval result re-ranking with LLMs.

Nevertheless, most state-of-the-art RAG systems still rely on chunk-based vector indexing,
and several graph-based indexing approaches are primarily tailored for knowledge base construction,
and thus are less suitable for document question answering tasks
\cite{guoLightRAGSimpleFast2025,edgeLocalGlobalGraph2025,maHetGPTHarnessingPower2024,prockoGraphRetrievalAugmentedGeneration2024,xuRetrievalAugmentedGenerationKnowledge2024}.
This omission limits their ability to exploit the original, well-defined document structure for retrieval,
thereby constraining retrieval effectiveness in DocQA tasks.

\section{Methodology}

\begin{figure*}[htbp]
  \centering
  \includegraphics[width=\textwidth]{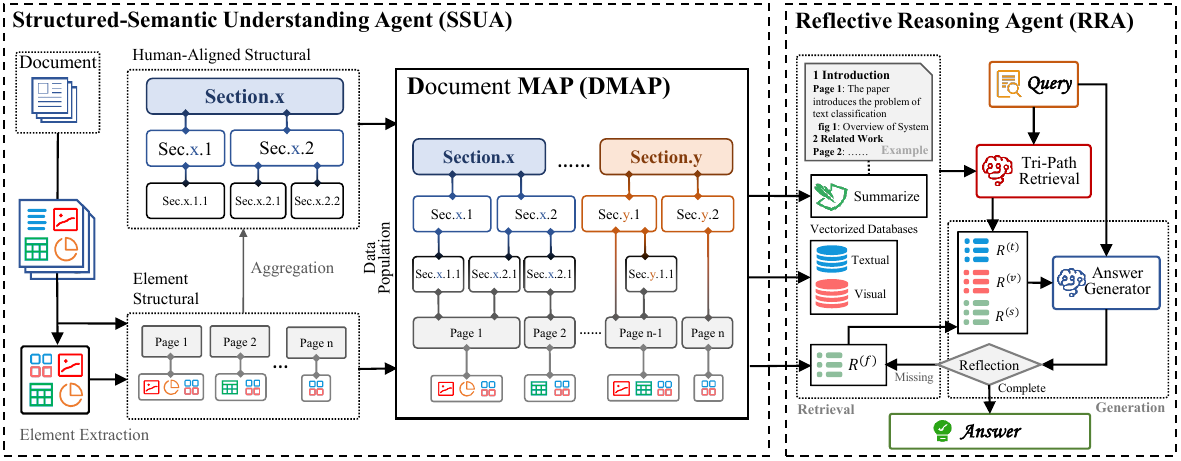}
  \Description{A workflow diagram illustrating the three-stage pipeline of the proposed method:
    offline processing, retrieval, and answer generation,
    with arrows showing the flow from document input to answer synthesis.}
  \caption{Overview of the framework. Structured-Semantic Understanding Agent (SSUA) constructs DMAP by organizing the document elements into a human-aligned hierarchical schema. Reflective Reasoning Agent (RRA) generates answers with DMAP in a reflective manner. }
  \label{fig:workflow}
\end{figure*}
In this section, we present our human-aligned framework for multimodal document question answering. Figure~\ref{fig:workflow} illustrates the overall framework that consists of two core agents: the Structured-Semantic Understanding Agent (SSUA) (the left part), which constructs DMAP, and the Reflective Reasoning Agent (RRA) (the right part), which leverages DMAP to perform structure-aware reasoning over multimodal documents. 

\subsection{Overall Framework}
Formally, given a multimodal document $D$ and a query $q$, our goal to generate the human-aligned structural document map $\mathcal{M}$ of $D$ using SSUA:
\begin{equation}
\mathcal{M} = \text{SSUA}(\mathcal{D}).
\end{equation}
Then, generate a semantically correct answer $A$ based on $\mathcal{M}$:
\begin{equation}
A = \text{RRA}(q \mid \mathcal{M}),
\end{equation}
where $\mathcal{M}$ encodes both hierarchical and cross-modal relationships among textual, tabular, and visual elements. 
This ensures that reasoning is aligned with the inherent structure of $D$, in contrast to conventional chunk-based RAG approaches.

\subsection{Structured-Semantic Understanding Agent (SSUA)}
The primary function of the Structured-Semantic Understanding Agent (SSUA) is to parse a raw multimodal document $\mathcal{D}$ into a human-aligned hierarchical structure (DMAP) and populate it with multimodal content and element associations, providing structured knowledge for downstream reasoning and question answering. As shown in the left part of Figure~\ref{fig:workflow}, DMAP is designed to reflect the hierarchical and relational organization of human-authored documents, making it intuitive for downstream reasoning and multimodal QA tasks. Formally, DMAP is a hierarchical prior capturing both the organization and multimodal elements of a document:

\begin{verbatim}
<document> ::= <section>+
<section> ::= <section title> ( <section>+ | <page>+ )
<page> ::= <element_set>
<element_set> ::= <element>*
<element> ::= ( figure | table | chart | page_content )
                <description> <location>
\end{verbatim}
This structure is designed to reflect the content logic in documents:

\begin{itemize}[leftmargin=*]
    \item \textbf{Section-level hierarchy:} Sections are organized according to the logical flow of the document, forming a tree-like structure that enables multi-level associations between content units. Sections serve as the backbone for capturing document-level organization.
    
    \item \textbf{Page-level concept:} A <page> represents the overall page as a logical unit in the document hierarchy. It contains the collection of all fine-grained elements on the page. The page thus serves as the primary anchor for locating content and linking higher-level sections to finer-grained elements, as well as for cross-page references.
    
    \item \textbf{Element-level representation:} Elements—figures, tables, charts, and page\_content (representing the full page with text and screenshot)—are the core units for fine-grained content comparison. Each element has both a visual signal (image or screenshot) and a textual description, supporting multimodal alignment, content reasoning, and QA tasks.
\end{itemize}

Based on the common logical structure of documents, we parse the content of each document and populate it into the hierarchical structure, thereby constructing a document-specific DMAP for downstream use.

\subsubsection{Element Extraction}
SSUA first decomposes the document $\mathcal{D}$ into a set of pages:
\begin{equation}
  \mathcal{D} = \{P_1, P_2, \dots, P_n\}.
\end{equation}

For each page $P_i$, SSUA extracts the key elements, which include:
\begin{itemize}
    \item \textbf{page\_content:} the full page, represented by its screenshot and textual content;
    \item \textbf{figures, tables, charts:} individual multimodal content items, each represented visually (image) and textually (caption or annotation).
\end{itemize}

The set of elements in page $P_i$ is denoted as:
\begin{equation}
  P_i = \{ e_{i1}, e_{i2}, \dots, e_{im} \}.
\end{equation}

For each element $e_{ij}$, we obtain abstracted multimodal descriptions using pretrained models: $v^T_{e_{ij}}$, $v^V_{e_{ij}}$,
which encode textual and visual characteristics, respectively.  
These descriptions will later support both content reasoning and similarity computation.

\subsubsection{Data Population}
After element extraction, SSUA populates DMAP structure with three main types of content and associations, reflecting the hierarchical organization of the document:

\paragraph{Elements:}  
Each page consists of a set of fine-grained elements, including figures, tables, charts, and the full-page content (`page\_content').  
These elements retain their textual annotations and visual representations, forming the atomic units for content comparison, reasoning, and multimodal QA.  
By explicitly modeling each element, DMAP captures detailed intra-page semantics while preserving multimodal alignment.

\paragraph{Pages:}  
Pages act as intermediate aggregation units that naturally organize their constituent elements into a coherent structure.  
A page $P_i$ contains all its elements, including `page\_content', and preserves the internal relationships among them.  
Pages serve as logical anchors linking element-level content to higher-level sections, supporting both intra-page reasoning and cross-page associations in DMAP hierarchy.

\paragraph{Sections:}
Using headings, layout cues, and semantic signals, elements and pages are hierarchically organized into sections, forming the backbone of the document’s structural map.
The construction of sections is a progressive process -- as each page is parsed, DMAP is gradually enriched until the complete structural representation emerges.
At step $i$, the current structural state of the document is represented as $S_i^\text{section}$, which encodes the accumulated hierarchy and relationships discovered so far:
\begin{equation}
S_i^\text{section} = \mathcal{A}_S^\text{section}(S_{i-1}^\text{section}, P_i, P_{i-1}),
\end{equation}
where $\mathcal{A}_S^\text{section}$ incrementally integrates the new page $P_i$ with the existing structural context $S_{i-1}^\text{section}$, updating section boundaries, hierarchies, and semantic dependencies.
Through this recursive refinement, DMAP progressively materializes into a coherent structural knowledge representation, capturing both local coherence within sections and the global organization of the entire document.

Through the aforementioned element extraction and data population processes,
DMAP gradually takes shape as the hierarchical structure is progressively enriched across pages and sections.
As shown in Figure~\ref{fig:workflow}, the resulting DMAP functions as the structured knowledge base that underpins the subsequent multimodal reasoning and answer generation.

\subsection{Reflective Reasoning Agent (RRA)}

After constructing the document-level repository $R_{\mathcal{D}}$ and the structured DMAP $M_{\mathcal{D}}$, 
the Reflective Reasoning Agent (RRA) performs multimodal document question answering 
under a Retrieval Augmented Generation (RAG) framework. 
It operates in two major stages: 
\textit{retrieval} and \textit{reflective generation}. 
The former identifies relevant evidence elements across modalities, 
while the latter synthesizes and refines answers through structured reflection.

\subsubsection{Tri-Path Retrieval over DMAP}

Given a query $q$, RRA retrieves relevant document elements (text, figure, table, chart, etc.) 
via three complementary retrieval paths, all grounded in DMAP structure. As shown in Figure~\ref{fig:workflow}, DMAP is converted to two visual and textual vectorized databases and summarizations for the following retrieval manners.

\paragraph{(1) Structured Semantic Retrieval.}  
A structure-aware agent $\mathcal{A}_S$ leverages the hierarchical sections, pages, and elements summary within DMAP $M_{\mathcal{D}}$ to locate semantically relevant elements.  
Specifically, it navigates the summary tree $M_{\mathcal{D}}^\text{summary}$, interprets the textual descriptions of sections and pages, and identifies elements whose content aligns with the query intent:
\begin{equation}
R^{(s)} = \mathcal{A}_S(q, M_{\mathcal{D}}^\text{summary}),
\end{equation}
where $R^{(s)}$ denotes the set of elements selected from DMAP based on their conceptual relevance to the query. 


By using the summary information rather than raw element content, $\mathcal{A}_S$ can efficiently focus on the elements that are most likely to contain relevant knowledge while respecting the document’s hierarchical structure.
\paragraph{(2) Textual Feature Retrieval.}
The query is encoded into a textual representation $\mathbf{v}_q^T = f_{\text{text}}(q)$.
Elements in $M_{\mathcal{D}}$ are represented by their textual embeddings $\mathbf{v}_e^T$, 
and the top-$k$ most similar elements are retrieved:
\begin{equation}
R^{(t)} = \operatorname{TopK}_{e \in M_{\mathcal{D}}}
\left( \operatorname{sim}(\mathbf{v}_q^T, \mathbf{v}_e^T) \right).
\end{equation}

\paragraph{(3) Visual Feature Retrieval.}
Similarly, for visual understanding, 
the query is projected into the visual feature space $\mathbf{v}_q^V = f_{\text{vis}}(q)$,
and visually related elements (figures, charts, tables) are retrieved:
\begin{equation}
R^{(v)} = \operatorname{TopK}_{e \in M_{\mathcal{D}}}
\left( \operatorname{sim}(\mathbf{v}_q^V, \mathbf{v}_e^V) \right).
\end{equation}

\paragraph{Fusion.}
The final retrieval result aggregates all three paths:
\begin{equation}
R = R^{(s)} \cup R^{(t)} \cup R^{(v)}.
\end{equation}

This tri-path retrieval integrates structural, textual, and visual cues,
allowing the agent to retrieve semantically rich and structurally grounded context for subsequent reasoning.
\subsubsection{Reflective Generation}

Given the aggregated retrieval result 
$R = \{ E_i \mid E_i \in \mathcal{M}_{\mathcal{D}} \}$,  
which contains multimodal elements (figures, tables, charts, and page\_content) identified during retrieval,  
the Reflective Generation module synthesizes the final answer through iterative reasoning and structured reflection over DMAP.

Each element $E_i$ in $R$ is processed to extract its multimodal features:  
\begin{equation}
\mathbf{e}_i^{T} = \mathrm{f}_{\text{text}}(E_i), \quad
\mathbf{e}_i^{V} = \mathrm{f}_{\text{vis}}(E_i),
\end{equation}
where $\mathbf{e}_i^{T}$ and $\mathbf{e}_i^{V}$ denote the textual and visual features, respectively.  
The sets of all textual features $\{\mathbf{e}_i^T\}$ and visual features $\{\mathbf{e}_i^V\}$ are then separately provided to the generative reasoning engine.

We then adopt MDocAgent~\cite{hanMDocAgentMultiModalMultiAgent2025} as generator $\mathrm{G}(\cdot)$ to generate the answer.
$\mathrm{G}(\cdot)$ takes the textual feature set and the visual feature set as input and generates candidate answers that integrate both modalities:
\begin{equation}
a = \mathrm{G}\big(q, \{\mathbf{e}_i^T\}, \{\mathbf{e}_i^V\}\big).
\end{equation}

To ensure semantic completeness and structural consistency, we devise a reflective assessment to the generated answer using an LLM-based evaluator:  
\begin{equation}
\text{done} = \mathcal{A}_{\text{eval}}(q, a).
\end{equation}
If `$\text{done} = \text{no}$', indicating that the answer is incomplete, the agent traverses DMAP hierarchy to retrieve additional related elements -- such as neighboring or parent elements -- and provides their textual and visual features back to $\mathrm{G}(\cdot)$ for iterative refinement. 

This reflective process effectively combines the “R” and “G” components of the RAG framework:  
the retrieval stage grounds reasoning in multimodal evidence,  
while the generative stage iteratively integrates textual and visual features to produce a coherent, complete, and structurally consistent answer aligned with DMAP representation.

\section{Experiment}
\subsection{Experiment Setup}
\begin{table}[H]
  \caption{Statistics of datasets.}
  \begin{tabular}{lcc}
    \toprule
    Dataset     & Document Count & Question Count \\
    \midrule
    MMLongBench & 134            & 1073           \\
    LongDocURL  & 396            & 2325           \\
    PaperTab    & 307            & 393            \\
    PaperText   & 1086           & 2798           \\
    FetaTab     & 871            & 1016           \\
    \bottomrule
  \end{tabular}
  \label{tab:dataset}
\end{table}
\subsubsection{Implementation Details}
For model selection, we adopt GPT-4o as the backbone model for both our proposed method and the RAG method baseline.
In addition, we also evaluate the performance when using Qwen-plus.
For vector representations, we employ ColBERTv2~\cite{santhanamColBERTv2EffectiveEfficient2022} as the textual embedding model
and ColPali~\cite{faysseColPaliEfficientDocument2024} as the visual embedding model.
Retrieval is performed using the respective ColBERTv2 and ColPali retrievers.
All embedding and retrieval processes are conducted on 2 NVIDIA 3090 GPUs.
For PDF document parsing, we utilize the \texttt{pymupdf} library in Python for content segmentation and extraction,
and the \texttt{pdffigure2} tool for extracting key elements of the document.
In configuring the number of retrieved passages for the RAG process, 
we experiment with two settings: $top\text{-}1$ and $top\text{-}4$. 
The retrieval limit is applied independently to each retrieval stream, 
thereby maintaining a balanced number of candidate contexts between the textual and visual modalities.

\subsubsection{Dataset}
To validate the effectiveness of our proposed method, we conducted experiments on five benchmark datasets.
An overview of these datasets is provided in Table~\ref{tab:dataset}.
MMLongBench~\cite{maMMLONGBENCHDOCBenchmarkingLongcontext2024}
is a dataset designed to comprehensively evaluate the capabilities of document understanding.
The answers to its questions depend on various types of information sources and locations.
In addition, approximately 20.6\% of the questions are unanswerable,
a deliberate design intended to assess whether DocQA systems exhibit hallucination problems.
LongDocURL~\cite{dengLongDocURLComprehensiveMultimodal2025}
is a multimodal dataset that targets long documents.
It contains a substantial number of cross-modal questions and aims to evaluate system performance in processing long texts.
PaperText and PaperTab~\cite{huiUDABenchmarkSuite2024}
are both datasets focused on scientific paper understanding.
They are partitioned according to document structure and information types,
where PaperText emphasizes textual content comprehension, while PaperTab focuses on the interpretation of table data.
FetaTab~\cite{huiUDABenchmarkSuite2024},
derived from FetaQA~\cite{nanFeTaQAFreeformTable2022},
is a Wikipedia-based QA dataset containing a large proportion of tabular and chart information.

\subsubsection{Baselines}
We selected two types of MMDocQA system architectures as baselines:
LVLM-based and RAG-based MMDocQA system.
In the pure LVLM-based approach, the document is directly provided as a context to the LVLM for question answering.
For the RAG-based approach, we adopt MDocAgent \cite{hanMDocAgentMultiModalMultiAgent2025} as the baseline and replace its backbone model with GPT-4o---the same model used in our method.
MDocAgent leverages multi-agent collaboration and a retrieval strategy primarily guided by semantic embeddings.

\subsubsection{Metrics}
For all datasets, we follow the evaluation protocols used in
MDocAgent,
LongDocURL, and
MMLongBench,
employing GPT-4o to evaluate the outputs.
Given a question and its reference answer,
GPT-4o compares the output of the MMDocQA system and returns a boolean judgment indicating whether the answer is both complete and correct.
The overall accuracy is used as the primary evaluation metric for the MMDocQA systems.

\subsection{Main Results}
\begin{table*}[htbp]
\caption{Comparison of our method with state-of-the-art (SOTA) MMDocQA approaches. Both pure LVLM-based and RAG-based methods are evaluated using accuracy (\%) as the metric.}
\centering
\begin{tabular}{llcccccc}
\toprule
\textbf{Category} & \textbf{Method} & \textbf{MMLongBench} & \textbf{LongDocURL} & \textbf{PaperTab} & \textbf{PaperText} & \textbf{FetaTab} & \textbf{Avg} \\
\midrule
\multirow{2}{*}{\textbf{LVLM-based}} 
& Qwen-2.5-VL-7B-Instruct & 0.204 & 0.398 & 0.156 & 0.203 & 0.350 & 0.274 \\
& LLaVA-1.6-7B-Instruct  & 0.176 & 0.110 & 0.102 & 0.178 & 0.301 & 0.173 \\
\midrule
\multirow{3}{*}{\textbf{RAG-based}} 
& MDocAgent (Top-4) & 0.338 & 0.561 & 0.347 & 0.536 & 0.640 & 0.484 \\
& Ours (Top-1) & 0.350 & 0.568 & 0.362 & 0.508 & 0.669 & 0.437 \\
& \textbf{Ours (Top-4)} & \textbf{0.432} & \textbf{0.607} & \textbf{0.390} & \textbf{0.567} & \textbf{0.725} & \textbf{0.544} \\
\midrule
& \textbf{Improvement (\%)} & 27.8\% & 8.2\% & 12.3\% & 5.8\% & 13.3\% & 12.4\% \\
\bottomrule
\end{tabular}
\label{tab:main_results}
\vspace{-2mm}
\end{table*}

\begin{table*}[htbp]
  \caption{Ablation study results. The first three columns indicate whether Structured Semantic Retrieval, Textual Feature Retrieval and Visual Feature Retrieval are used in the configuration.}
  \begin{tabular}{c|ccc|ccccccc}
    \toprule
    \multirow{2}*{Variants} & \multicolumn{3}{c|}{Configuration} & \multicolumn{6}{c}{Evaluation Benchmarks} \\
    \cline{2-10}
    ~           & Text       & Image     & DMAP    & MMLongBench    & LongDocURL  & PaperTab   & PaperText   & FetaTab        & Avg. Acc       \\
    \cline{1-10}
    w/o DMAP & \ding{51}  & \ding{51} & \ding{55}  & 0.320    & 0.534      & 0.337          & 0.531          & 0.514          & 0.447  \\
    w/o Image   & \ding{51}  & \ding{55} & \ding{51}  & 0.380    & 0.593      & 0.385          & 0.565          & 0.705          & 0.526  \\
    w/o Text    & \ding{55}  & \ding{51} & \ding{51}  & 0.390    & 0.482      & 0.219          & 0.223          & 0.659          & 0.394  \\
    Full        & \ding{51}  & \ding{51} & \ding{51}  & \textbf{0.432} & \textbf{0.607} & \textbf{0.390} & \textbf{0.567} & \textbf{0.725} & \textbf{0.544} \\
    \bottomrule
  \end{tabular}
  \label{tab:ablation}
  \vspace{0 mm}
\end{table*}

The comparative results between our method and various baselines are presented in Table~\ref{tab:main_results}.
As shown in Table~\ref{tab:main_results}, our method achieves the best performance across all benchmarks, demonstrating its strong ability in MMDocQA tasks. 
Specifically, Ours(Top-4) outperforms the strongest baseline (MDocAgent) by a notable margin, achieving an average accuracy improvement of 12.4\%.

The most significant improvement is observed on MMLongBench, 
a dataset characterized by a high proportion of complex questions, 
such as those requiring multi-page reasoning and comprehension of elements.
Benefiting from the section-level hierarchy and the structured retrieval via DMAP, 
our method excels at integrating information from multiple pages and organizing it effectively, 
thereby overcoming a key limitation of conventional RAG methods in handling cross-page DocQA.
Moreover, our fine-grained analysis and localization of critical elements enhance performance in questions involving complex reasoning over figures and tables.
This capability is further reflected in the substantial gains obtained on PaperTab and FetaTab, 
both of which contain a large amount of tabular and graphical data and are designed with questions closely tied to multimodal understanding.

In contrast, PaperText primarily consist of textual content with relatively few diagrams or tables, 
making it more challenging to fully leverage the multimodal capabilities of our approach. 
Nevertheless, our method still achieves competitive improvements on the dataset, 
suggesting that the section-level hierarchy offers advantages even in mainly text-based scenarios.

In terms of the number of candidate contexts, 
the performance improvement achieved by Top-4 retrieval is consistently greater than that achieved by Top-1 retrieval, 
indicating that the amount of supporting evidence plays a crucial role in MMDocQA performance.
This effect is particularly pronounced in datasets such as MMLongBench, where many questions require multi page reasoning. 
Nevertheless, even with Top-1 retrieval, our method occasionally outperforms the MDocAgent (Top-4) baseline, 
highlighting the effectiveness of our structured retrieval, 
which enables a more precise definition of contextual scope and allows our method to achieve better results with fewer context segments.

When comparing different architectures for MMDocQA, 
all RAG-based method exhibit substantial performance improvements across all benchmarks, highlighting the the necessity of integrating RAG techniques in MMDocQA tasks.

Overall, these findings validate the effectiveness of our proposed framework in enhancing RAG for multimodal document understanding,
particularly in scenarios involving high reasoning complexity, multi-page dependencies, and multimodal contexts.

\subsection{Ablation Study}
\subsubsection{Contribution of Different Retrieval Paths.} 
To assess the effectiveness of the core retrieval paths in our framework, we conducted an ablation study focusing on the three retrieval paths:
Structured Semantic Retrieval, Textual Feature Retrieval and Visual Feature Retrieval.
We systematically removed each path and evaluated performance across multiple DocQA benchmarks.
The results are summarized in Table~\ref{tab:ablation}.
As shown in the table, removing any single module leads to a noticeable performance drop,
demonstrating both the necessity of each component and their synergistic contributions to the overall framework.

In particular, The largest average drop occurs when removing the Textual Feature Retrieval, resulting in a $-15.0\%$ decrease in average accuracy. 
This is primarily because the PaperTab and PaperText datasets are predominantly text-based; 
therefore, excluding the textual retrieval module severely degrades performance on these benchmarks. 
This observation highlights that mainstream text retrieval approaches still demonstrate strong effectiveness when handling text-dominant modalities.

Removing the Structured Semantic Retrieval module (Summary) leads to the second largest performance degradation ($-9.7\%$), 
especially on MMLongBench and FetaTab. 
Specifically, MMLongBench involves extensive multi-page reasoning, complex problem-solving, and multimodal question-answering tasks 
and FetaTab emphasizes the comprehension of multimodal tabular information. 
These observations demonstrate that incorporating the Structured Semantic Retrieval significantly enhances performance on tasks requiring complex reasoning and multimodal understanding.

Similarly, removing the Visual Feature Retrieval module decreases the average performance to $0.526$ ($-1.8\%$). The relatively small drop in the overall mean score can be primarily attributed to the fact that this module provides limited benefits on the two text-centric DocQA datasets, PaperTab and PaperText, compared with the other retrieval modes.  
This observation suggests that although current LVLMs possess basic OCR capabilities and can process certain textual information through the visual modality, 
their ability to handle purely textual information remains inferior to that of retrieval pipelines leveraging the textual modality alone. 
Nonetheless, image-based retrieval and generation serve as a crucial complement to textual retrieval. 
Therefore, integrating both modalities is the most effective approach.


\subsubsection{Effect of the Reflection Module.} 
To evaluate the effectiveness of the reflection module, we conducted a standalone experiment. 
We analyze the necessity of the reflection mechanism across different datasets.
For a fair comparison across datasets, we randomly sampled 1,000 questions for evaluation,
which we consider sufficiently large to be representative of the overall dataset distribution.
The results are listed in Table~\ref{tab:reflection}. 
\begin{table}[ht]
  \caption{Impact of the reflection module.}
  \centering
  \begin{tabular}{lccc}
    \toprule
    Dataset     & Correct / Error & Count & Accuracy \\
    \midrule
    MMLongBench & 185 / 96        & 281   & 0.658    \\
    FetaTab     & 36 / 30         & 66    & 0.545    \\
    LongDocURL  & 90 / 81         & 171   & 0.526    \\
    PaperText   & 56 / 14         & 70    & 0.800    \\
    \bottomrule
  \end{tabular}
  \label{tab:reflection}
\end{table}
The PaperTab dataset was excluded from this analysis due to its limited number of available test questions.
Notably, the MMLongBench dataset required the largest number of reflection operations, as 20.6\% of its questions are designed to be unanswerable.
DocQA systems are often unable to determine whether a question is unanswerable due to its deliberate question design or the insufficiency of the retrieved context,
in such cases, the reflection mechanism is subsequently activated.
However, our reflection module demonstrates strong \textbf{Negative Rejection} capabilities:
when confronted with these difficult, unanswerable questions, it is able to faithfully respond `unknown' rather than exhibit the \textbf{hallucination} phenomenon
\cite{huangSurveyHallucinationLarge2025,gaoRetrievalAugmentedGenerationLarge2024}.
This behavior is a key factor in introducing RAG systems into QA tasks, and our method effectively leverages this advantage.
For the FetaTab and LongDocURL datasets, the reflection mechanism also contributed
significantly to performance stability.
While slight accuracy drops are observed, such decreases are anticipated,
as the reflection mechanism primarily targets challenging cases,
specifically whose answer was already identified by the DocQA system as incorrect or incomplete.

Overall, the results in Table~\ref{tab:reflection} affirm that the reflection module not only
improves the system's ability to handle complex questions but also
exhibits strong robustness when confronted with tasks that require negative rejection.

\subsection{Mechanism Analysis of DMAP}

\subsubsection{Evidence Source Analysis:}
To understand the underlying reasons for the improvements achieved by our method,
we conducted a detailed analysis based on the evidence sources required to answer each question.
In the MMLongBench dataset, questions are categorized according to the type of evidence they rely on,
falling into the following five categories:
(1) Pure Text, (2) Layout, (3) Table, (4) Chart, and (5) Figure.
Specifically, Pure Text refers to questions whose answers can be fully recovered from the textual content;
Layout indicates that correctly answering the question requires reasoning over the structural layout;
Table, Chart, and Figure correspond to questions requiring information extraction from tabular data, charts, and figures.

Table~\ref{tab:source} presents the number of questions in each category, and result of the experiment.Numbers in parentheses denote the number of samples in each category.Some questions require multiple evidence sources; thus, the total count across categories exceeds the total number of unique questions in the dataset.
Our method achieves consistent improvements across all evidence sources,
with particularly substantial gains observed in the Chart and Figure categories
(accuracy improvements of $+89.4\%$ and $+74.5\%$, respectively).
This aligns with our earlier speculation that our method is especially effective in handling visual information.
In contrast, improvements in the Table category are more modest ($+16.8\%$),
which can be attributed to two factors:
(i) the document parsing tools used in our pipeline already preserve tabular structure with high fidelity,
and (ii) GPT-4o itself possesses strong inherent table comprehension capabilities.
As a result, the improvement in table-based questions is comparable to that of pure textual questions.
For Layout-based questions, our method achieves a remarkable $+47.9\%$ improvement,
demonstrating that the structured document indexing strategy significantly enhances the model's ability to interpret
overall document organization and layout semantics.

\begin{table}[ht]
  \caption{Accuracy on different evidence source categories in MMLongBench.}
  \begin{tabular}{lccc}
    \toprule
    Evidence Source  & Ours & MDocAgent & Improvement \\
    \midrule
    Pure Text  (281) & \textbf{0.470}     & 0.398     & 18.1\%      \\
    Layout (118)     & \textbf{0.339}     & 0.229     & 47.9\%      \\
    Table (162)      & \textbf{0.432}     & 0.370     & 16.8\%      \\
    Chart (175)      & \textbf{0.377}     & 0.199     & 89.4\%      \\
    Figure (299)     & \textbf{0.349}     & 0.200     & 74.5\%      \\
    \bottomrule
  \end{tabular}
  \label{tab:source}
\end{table}

\begin{table*}[htbp]
  \caption{Performance across different LVLM backbones.}
  \resizebox{0.9\linewidth}{!}{
  \begin{tabular}{lcccccc}
    \toprule
    Model                                & MMLongBench    & LongDocURL     & PaperText      & PaperTab       & Feta Tab       & Avg. Acc       \\
    \midrule
    Qwen-2.5-VL-7B-Instruct              & 0.319          & 0.300          & 0.476          & 0.286          & 0.625          & 0.401          \\
    Qwen-plus \& Qwen-vl-plus(MDocAgent) & 0.331          & 0.502          & 0.495          & 0.242          & 0.609          & 0.435          \\
    Qwen-plus \& Qwen-vl-plus            & 0.353          & 0.555          & 0.519          & 0.368          & 0.675          & 0.494          \\
    GPT-4o                               & \textbf{0.432} & \textbf{0.607} & \textbf{0.567} & \textbf{0.390} & \textbf{0.725} & \textbf{0.544} \\
    \bottomrule
  \end{tabular}
  \label{tab:diff_model}
  }
  \vspace{0 mm}
\end{table*}

While all categories benefit from our approach, the large gains in visual evidence sources (Chart and Figure)
suggest that conventional baselines underutilized such modalities,
possibly due to limitations in their multimodal fusion mechanisms or insufficient training on visually grounded reasoning tasks.
The DMAP not only preserves spatial and relational cues from these visual elements but also facilitates better integration with textual context.
\begin{figure*}[ht]
  \centering
  \includegraphics[width=0.85\textwidth]{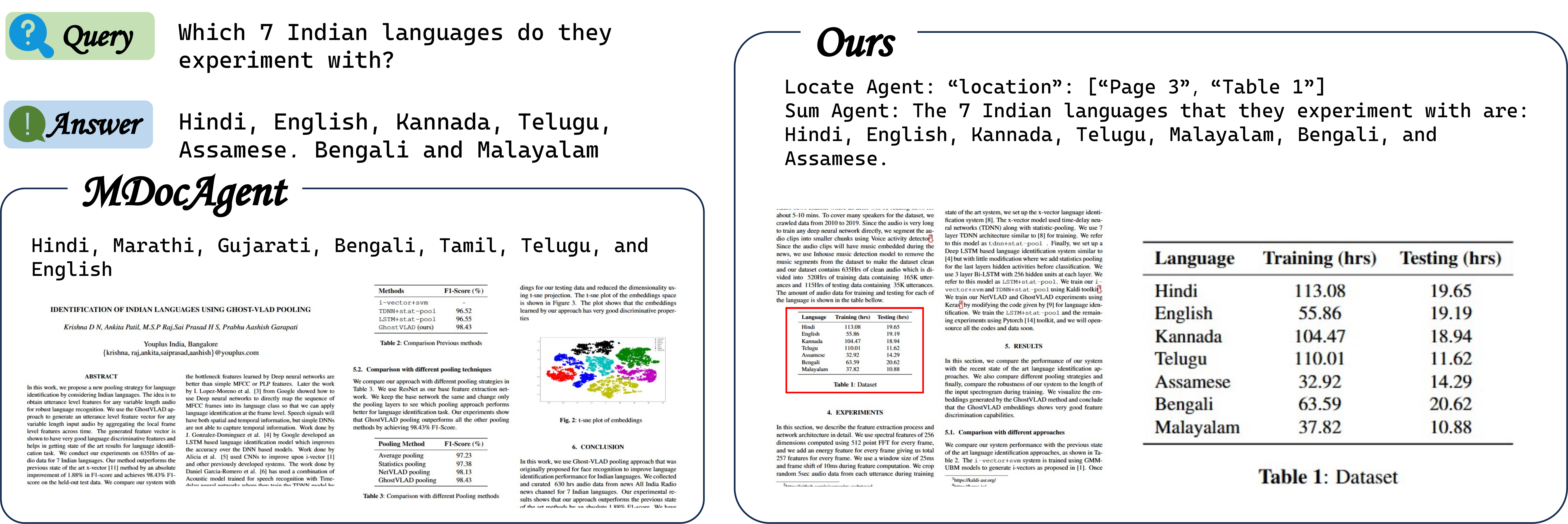}
  \Description{Case Study}
  \caption{
    Qualitative comparison between MDocAgent and our method on a table-centric DocQA task.
    The question requires integrating information from a tabular dataset and text descriptions.
    Our method accurately identifies and parses the relevant table to produce the correct answer.
  }
  \label{fig:case_study}
  \vspace{0 mm}
\end{figure*}
\subsubsection{Model Analysis:}
We further conduct experiments focused on the robustness and scalability of our method by evaluating its performance across different LVLMs,
ranging from lightweight models to more powerful, large-scale models.
The results are presented in Table~\ref{tab:diff_model}.

Overall, stronger models yield a clear boost in performance across all benchmarks.
This improvement is particularly notable on MMLongBench,
demonstrating that to handle complex, multimodal tasks, comprehensive understanding ability is required.
A similar upward trend can be observed in LongDocURL,
where more capable models demonstrate improved reasoning consistency when processing multi-page contexts.

Importantly, our DocQA framework exhibits consistent gains over the baseline across all model sizes,
demonstrating its ability to extract and integrate multimodal cues effectively regardless of the underlying model capacity.
This suggests that our architecture is adept at exploiting the representational power of stronger LVLMs without requiring specific tuning.

Interestingly, even with a smaller model such as Qwen-2.5-VL-7B-Instruct,
our approach maintains competitive performance and, in some datasets (PaperTab and FetaTab), exceeds the baseline of much larger models.
This is likely due to these datasets having relatively short document lengths and less complex queries.
In such scenarios, the limitations of smaller LVLMs in complex queries are less exhibited,
allowing our method to achieve performance gains over model ability.

These results confirm two desirable properties of our method:
a) scalability, in that performance grows consistently with stronger LVLMs,
and b) robustness, in that its effectiveness does not heavily degrade when applied to smaller models.
This indicates that our method is a practical and future-proof solution for MMDocQA tasks,
capable of adapting to varying resource budgets while reaping the benefits of advancements in LVLM architectures.
\subsection{Case Study}
To gain deeper insights into mechanisms of our method,
we conducted a qualitative case study, as illustrated in Figure~\ref{fig:case_study}.
The posed query is: `Which 7 Indian languages do they experiment with?'
In the document, the descriptions of the experimented languages are only partially provided in the main text,
while the complete information is provided in the table in the experimental section.
This presents three specific challenges for a DocQA system:
(1) recognize that the primary evidence source for the query lies in a table rather than in main text,
(2) accurately locate the relevant table in a multi-page document,
(3) understand information in table while avoiding the noise from incomplete textual information.

In this case, the baseline model MDocAgent retrieves pages
most relevant to `experiment' by semantic relevance between the query and these pages.
However, due to the document layout, the critical table does not in the same page.
This makes MDocAgent partially misinterpreting the data and producing an incorrect answer,
which omits some correct languages and includes irrelevant ones.

In contrast, DMAP employs a fine-grained layout-aware document parsing mechanism to detect Table~\ref{tab:dataset},
thus overcoming the page retrieval limitations.
Once located, our method accurately extracts the correct data from the tabular content:
Hindi, English, Kannada, Telugu, Assamese, Bengali, and Malayalam.

In summary, the case study illustrates that our structured document understanding framework can
pinpoint and parse key document elements reliably,
resulting in performance gains for DocQA tasks,
particularly in queries where the answer’s evidence source depends on multi-source information.
\section{Conclusion}
In this work, we presented DMAP, a human-aligned structural representation designed to capture the hierarchical and relational organization of multimodal documents.
Built upon this representation, we developed a Structured-Semantic Understanding Agent (SSUA) for structure-aware retrieval and a Reflective Reasoning Agent (RRA) for iterative answer synthesis, together forming a unified framework that aligns machine reasoning with human comprehension.
By explicitly modeling the interplay among textual, tabular, and visual elements, our approach moves beyond flat semantic retrieval and demonstrates a scalable path toward structure-informed multimodal understanding.

In future, we will extend DMAP toward dynamic, self-evolving document cognition, enabling agents to incrementally construct, refine, and reason over multimodal knowledge at scale.
This direction paves the way for next-generation document intelligence systems—models capable of not only extracting and generating information, but also engaging in reflective, context-aware reasoning across complex and evolving document ecosystems.
\clearpage
\bibliographystyle{ACM-Reference-Format}
\bibliography{reference}
\clearpage
\appendix
\section{Prompt}
To better illustrate how each of our agents operates, below are the example prompts we use for every agent.
\subsection{Locate Agent Prompt}
\begin{tcolorbox}[
  width=\columnwidth,     
  breakable,  
  colback  = gray!15,     
  colframe = gray!50,     
  arc      = 4pt,         
  boxrule  = 0.6pt,       
  left     = 6pt, right = 6pt, top = 4pt, bottom = 4pt  
]
You are an expert assistant for analyzing questions in a RAG system. Your task is to identify and infer the most relevant locations in a document based on the question, document summary, and outline.\\

\# Task: Location Identification, Inference, and Ranking\\
I will provide:\\
1. A document summary, summarized page by page.\\
2. A document outline in the format: "\{section number\} \{section name\} <|> pages"\\
3. A question that may refer to specific content.\\

Your task is to:\\
1. Extract explicit location references from the question:\\
   - If a page number is mentioned (e.g., "Page 1", "page ii"), convert it to: `"Page \{number\}"`, where `\{number\}` is the integer form (e.g., "ii" → 2).\\
   - If a table or figure is mentioned, extract the number and format as: `"Table \{number\}"`, `"Figure \{number\}"`.\\
2. Infer implicit locations when no explicit reference exists:\\
   - Use the document summary and outline to determine the most likely page(s) related to the question.\\
   - Match keywords, topics, or concepts in the question to the summaries and section titles.\\
   - If a section in the outline covers the topic and lists specific pages, infer one or more of those pages as likely locations.\\
   - Always provide at least one plausible page if the topic can be reasonably located.\\
3. Rank all identified and inferred locations by relevance:\\
   - Sort the final list in descending order of relevance to the question.\\
4. If no location can be inferred at all, return `["not mentioned"]`.\\
Output all location references in a JSON array under the key `"location"`, ordered by relevance.\\

\# Output Format\\
Return your result in the following JSON format:\\
\{"location": ["Page 7", "Table 3", "Page 8", ...] | ["not mentioned"]\}\\

\# Input:\\
\end{tcolorbox}

\subsection{Summarize Agent Prompt}
\begin{tcolorbox}[
  width=\columnwidth,     
  breakable,  
  colback  = gray!15,     
  colframe = gray!50,
  arc      = 4pt, 
  boxrule  = 0.6pt,
  left     = 6pt, right = 6pt, top = 4pt, bottom = 4pt 
]
You are an expert assistant for analyzing questions in a RAG system. \\
\# Task: You are tasked with processing a document page by page to simultaneously:\\
1. Generate a concise summary of the current page.\\
2. Maintain and update a hierarchical document outline that maps section titles to the pages they span.\\

You will be given:\\
The current cumulative outline (as a numbered hierarchy with page mappings),\\
The previous page (for context),\\
The current new page to process,\\
The current page number.\\

For each new page, perform the following steps:\\
Step 1: Update the Document Outline\\
Analyze the current page for any explicit section headings (e.g., "1. Introduction", "3.2 Experimental Setup") or implicit topic shifts that suggest a new logical section.\\
If a new section heading is detected:\\
Assign it the next appropriate hierarchical number (e.g., if the last top-level was "2", a new top-level becomes "3"; if under "3", a new subsection becomes "3.1", etc.).\\
Add this section to the outline with its title and initialize its page list with the current page number.\\
If the content continues an existing section (even without a heading), identify the most specific (deepest-level) active section(s) that this page belongs to.\\
Important: A single page may belong to multiple sections (e.g., spanning the end of one subsection and the start of another). In such cases, add the page number to all relevant sections, including their parent sections.\\
Always propagate the page number upward to all ancestor sections in the hierarchy.\\
Step 2: Summarize the Current Page\\
1. Provide a one-sentence summary of the page's main content. If the page is blank or has no meaningful content, summarize it as "no content".\\
2. Identify any figures or tables on the page. For each:\\
If a caption or name is explicitly provided (e.g., "Figure 1: System Architecture"), use it exactly as written. If no label is given but a caption exists, format as Figure : [caption] or Table : [caption]. Do not invent names.\\
Provide a one-sentence description of the figure or table.\\
If there are no figures or tables, omit this part—only include it when present.\\
\# Output Format\\
Your response must strictly follow this structure:\\

Outline:\\
\{updated current full outline\}\\
\{each line formatted as: \{number\}:\{title\} < > \{comma-separated page numbers\}\}\\

Current page summary:\\
Page \{number\}: [One-sentence summary of the page content.]\\
Figure \{name\}: [One-sentence summary of the figure.]\\
Table \{name\}: [One-sentence summary of the table.]\\

OR, if the page has no content and no figures/tables:\\
Outline:\\
\{updated current full outline\}\\

Current page summary:\\
Page \{number\}: no content\\
no figure or table\\

Important Notes:\\
Only output the updated outline and the summary for the current page—do not repeat prior page summaries.Maintain consistent numbering and hierarchy in the outline across pages.Never fabricate section titles; infer only when strongly supported by content structure or semantic shift.Page numbers in the outline must be sorted and deduplicated.\\

\# Input:\\
\end{tcolorbox}

\subsection{Generator's Text Agent Prompt}
\begin{tcolorbox}[
  width=\columnwidth,     
  breakable,  
  colback  = gray!15,     
  colframe = gray!50,
  arc      = 4pt, 
  boxrule  = 0.6pt,
  left     = 6pt, right = 6pt, top = 4pt, bottom = 4pt 
]
You are a text analysis agent. Your job is to extract key information from the text and use it to answer the user’s question accurately. Here are the steps to follow:\\
Extract key details: Focus on the most important facts, data, or ideas related to the question.\\
Understand the context: Pay attention to the meaning and details.\\
Provide a clear answer: Use the extracted information to give a concise and relevant response to user's question.\\
Remember you can only get the information from the text provided, so maybe other agents can help you with the image information.\\
If the provided reference content cannot answer the question, do not add any extra explanation, directly output "not answerable".\\
Question:\\

\end{tcolorbox}

\subsection{Generator's Image Agent Prompt}
\begin{tcolorbox}[
  width=\columnwidth,     
  breakable,  
  colback  = gray!15,     
  colframe = gray!50,
  arc      = 4pt, 
  boxrule  = 0.6pt,
  left     = 6pt, right = 6pt, top = 4pt, bottom = 4pt 
]
You are an advanced image processing agent specialized in analyzing and extracting information from images. The images may include document screenshots, illustrations, or photographs. Your primary tasks include:\\
Extracting textual information from images using Optical Character Recognition (OCR).\\
Analyzing visual content to identify relevant details (e.g., objects, patterns, scenes).\\
Combining textual and visual information to provide an accurate and context-aware answer to user's question.\\
Remember you can only get the information from the images provided, so maybe other agents can help you with the text information.\\
If the provided reference content cannot answer the question, do not add any extra explanation, directly output "not answerable".\\
Question:\\

\end{tcolorbox}

\subsection{Generator's Summarize Agent Prompt}
\begin{tcolorbox}[
  width=\columnwidth,     
  breakable,  
  colback  = gray!15,     
  colframe = gray!50,
  arc      = 4pt, 
  boxrule  = 0.6pt,
  left     = 6pt, right = 6pt, top = 4pt, bottom = 4pt 
]
You are tasked with summarizing and evaluating the collective responses provided by multiple agents. You have access to the following information: \\
Answers: The individual answers from all agents.\\
Using this information, perform the following tasks:\\
1. Filter: Ignore any agent who indicates that they are unable or unwilling to provide an answer. Only consider responses from agents who explicitly offer a solution or reasoning.\\
2. Analyze: Evaluate the quality, consistency, and relevance of each valid answer. Identify commonalities, discrepancies, or gaps in reasoning.  \\
3. Synthesize: Summarize the most accurate and reliable information based on the evidence provided by the agents and their discussions.  \\
4. Conclude: Provide a final, well-reasoned answer to the question or task. Your conclusion should reflect the consensus (if one exists) or the most credible and well-supported answer.\\
Based on the provided answers from all agents, summarize the final decision clearly. You should only return the final answer in this dictionary format: \\
```json\\
\{"Answer": "<Your final answer here>"\}\\
```\\
Do not include any additional information or explanation.\\
\end{tcolorbox}

\subsection{Reflect Agent Prompt}
\begin{tcolorbox}[
  width=\columnwidth,     
  breakable,  
  colback  = gray!15,     
  colframe = gray!50,
  arc      = 4pt, 
  boxrule  = 0.6pt,
  left     = 6pt, right = 6pt, top = 4pt, bottom = 4pt 
]
You will be given a question and a corresponding answer. Your task is to determine whether the answer addresses the question, regardless of whether the answer is correct or not.\\

Focus only on whether the answer responds to the question and covers the necessary points (i.e., no essential content is missing).\\
If no answer is provided (e.g., blank, "not answerable", or similar), consider it as not answering.\\
Respond only with "yes" or "no", in lowercase. Do not include any explanations, punctuation, or additional text.\\
Question:\\
\{question\}\\
Answer:\\
\{answer\}\\
Did the answer address the question? (yes/no)\\
\end{tcolorbox}

\end{document}